\documentclass[prd,aps,lengthcheck,twocolumn,nofootinbib,notitlepage,floatfix,superscriptaddress]{revtex4-2}
\usepackage{graphics,graphicx}
\usepackage{amsmath, amssymb}
\usepackage{multirow}
\usepackage{color}
\usepackage{verbatim}
\usepackage{hyperref}
\usepackage[normalem]{ulem}
\usepackage{color}
\usepackage{cancel}
\usepackage{mathtools}
\usepackage{orcidlink}

\def\fm3{\,\text{fm}^{-3}}

\renewcommand\sout{\bgroup\color{blue} \ULdepth=-.5ex \ULset}
\def\slashchar#1{\setbox0=\hbox{$#1$}  % set a box for #1
\dimen0=\wd0     % and get its size
\setbox1=\hbox{/} \dimen1=\wd1  % get size of /
\ifdim\dimen0>\dimen1   % #1 is bigger
\rlap{\hbox to \dimen0{\hfil/\hfil}} % so center / in box
#1     % and print #1
\else     % / is bigger
\rlap{\hbox to \dimen1{\hfil$#1$\hfil}} % so center #1
/      % and print /
\fi}

\newcommand{\dd}{\mathrm{d}}
\newcommand{\pp}{\partial}

\allowdisplaybreaks

\hypersetup{
pdftitle={Quark-diquark structure of baryons from fluctuations},
pdfauthor={MM}
}

\begin{document}
\title{Evidence for quark-diquark structure of baryons from fluctuations of conserved charges}

\author{Micha\l{} Marczenko \orcidlink{0000-0003-2815-0564}}
\email{michal.marczenko@uwr.edu.pl}
\affiliation{Institute of Theoretical Physics, University of Wroc\l{}aw, plac Maksa Borna 9, PL-50204 Wroc\l{}aw, Poland}
\affiliation{International Institute for Sustainability with Knotted Chiral Meta Matter (WPI-SKCM$^2$), Hiroshima University, 1-3-1 Kagamiyama, Higashi-Hiroshima, Hiroshima 739-8531, Japan}
\author{Krzysztof Redlich \orcidlink{0000-0002-2629-1710}}
\email{krzysztof.redlich@uwr.edu.pl}
\affiliation{Institute of Theoretical Physics, University of Wroc\l{}aw, plac Maksa Borna 9, PL-50204 Wroc\l{}aw, Poland}
\affiliation{Polish Academy of Sciences PAN, Podwale 75, 
PL-50449 Wroc\l{}aw, Poland}
% %
 
\date{\today}

\begin{abstract}
We study fluctuations and correlations of conserved charges in QCD using a string-based description of the hadronic mass spectrum. Mesons and baryons are modeled as open relativistic strings with quark-antiquark and quark-diquark endpoints, respectively, leading to an exponential Hagedorn growth of states with a limiting temperature fixed by the string tension. We find that continuous Hagedorn spectra constrained by experimentally established hadrons underestimate net-baryon number fluctuations obtained in lattice QCD calculations. By extracting the Hagedorn string spectrum directly from lattice QCD through a fit to the second-order net-baryon number susceptibility, we obtain a consistent description of a broad set of fluctuations of conserved charges from LQCD with the Hagedorn temperature $T_H \simeq 323~$MeV, without introducing additional free parameters. Our results provide thermodynamic evidence in support of a string quark-diquark picture of baryons in the confined phase of QCD.
\end{abstract}

\maketitle

\section{Introduction}

Fluctuations and correlations of conserved charges provide a quantitative probe of {critical behavior associated with the existence of a putative critical point in the phase diagram of quantum chromodynamics (QCD)~\cite{Stephanov:1999zu, Asakawa:2000wh, Jeon:2000wg, Jeon:2000wg, Hatta:2003wn, Majumder:2006nq, Stephanov:2011pb, Friman:2011pf}, as well as color deconfinement and remnants of $O(4)$ chiral criticality~\cite{Friman:2011pf,Ejiri:2005wq, Bazavov:2011nk, Karsch:2019mbv, Braun-Munzinger:2020jbk, Braun-Munzinger:2016yjz, Tohme:2025nzw, Mitra:2024mke}. Fluctuations are also sensitive probes of} the degrees of freedom in QCD matter. High-precision lattice QCD (LQCD) calculations at finite temperature have shown that the susceptibilities of the net-baryon number, strangeness, and electric charge encode detailed information about the hadronic spectrum below the chiral crossover temperature \mbox{$T_c\simeq 155~$MeV}, as well as about the onset of deconfinement~\cite{Bazavov:2014pvz, Borsanyi:2018grb, HotQCD:2018pds, Bellwied:2019pxh, Bollweg:2021vqf}.

In the confined phase, thermodynamic observables are commonly described using the hadron resonance gas (HRG) model~\cite{Braun-Munzinger:2003pwq, Andronic:2017pug}, in which hadrons and resonances are effectively contributing to the thermodynamic potential as a mixture of ideal gases. In the spirit of the S-matrix approach to QCD thermodynamics, this corresponds to incorporating the leading attractive interaction contribution to the grand canonical potential through the virial expansion~\cite{Venugopalan:1992hy}.
While this approach successfully reproduces bulk thermodynamics of the hadronic phase, systematic deviations from LQCD results are observed in higher-order fluctuations around the chiral crossover temperature~\cite{Bazavov:2014xya, Lo:2015cca, ManLo:2016pgd, Bollweg:2021vqf}.

The rapid growth of the hadronic spectrum has long been associated with the concept of a Hagedorn mass spectrum, characterized by an exponential density of states and a limiting temperature for hadronic matter~\cite{Hagedorn:1965st, Hagedorn:1971mc}. Beyond its original formulation within the statistical bootstrap model, an exponential spectrum arises naturally in relativistic string theory, where the proliferation of string excitation modes leads to a universal Hagedorn behavior~\cite{Frautschi:1971ij, Green_Schwarz_Witten_2012}. In this framework, the Hagedorn temperature is fixed by the string tension, while the dimensionality and topology of the string determine the functional form of the spectrum.

Recently, string-inspired descriptions of hadronic spectra have been revisited in the context of QCD thermodynamics. Mesons and glueballs have been modeled as open and closed relativistic strings, respectively, leading to a unified description of their spectra and to a Hagedorn temperature of order $300\,$MeV, comparable to the deconfinement temperature of pure gauge theory~\cite{Fujimoto:2025sxx, Marczenko:2025nhj}. In this context, it was proposed that QCD may exhibit an intermediate regime between the hadronic phase and the quark-gluon plasma, in which quarks appear as effective thermal degrees of freedom while gluons remain confined in the form of glueballs~\cite{Fujimoto:2025sxx}. Likewise, a string-based description of baryons has been developed, in which baryons are modeled as open relativistic strings with a quark attached to one endpoint and a diquark attached to the other~\cite{Fujimoto:2025trl}. In this picture, the string dynamics governing baryons is identical to that of mesons, while baryon-specific properties enter only through endpoint quantum numbers and physical threshold masses. This construction naturally leads to an exponential baryonic spectrum governed by the same Hagedorn temperature as the mesonic sector, consistent with general expectations from relativistic string models.

In this work, we apply the string-based description of hadronic spectra to fluctuations and correlations of conserved charges, i.e., net-baryon number, strangeness, and electric charge. Our main observation is that fluctuations of conserved charges provide a particularly stringent test of the effective hadronic spectrum relevant for thermodynamics. Modeling such observables with continuous Hagedorn spectra constrained solely by hadronic spectroscopy is found to differ from the LQCD results. We therefore determine the effective Hagedorn spectrum directly from LQCD. The resulting spectrum yields parameter-free predictions for a broad set of conserved-charge susceptibilities and correlations and shows significantly improved agreement with LQCD. Our analysis provides thermodynamic evidence for a string-based hadronic mass spectrum, in particular for a quark-diquark structure of baryons.

This paper is organized as follows. In Sec.~\ref{sec:had_spectrum}, we summarize the discrete hadronic spectrum and its cumulative properties. In Sec.~\ref{sec:hagedorn}, we introduce the string-based Hagedorn mass spectrum for mesons and baryons, and extract the Hagedorn temperature from different inputs. Section~\ref{sec:fluct} is devoted to fluctuations and correlations of conserved charges and to the determination of the effective spectrum from lattice QCD. Our findings are summarized in Sec.~\ref{sec:conclusions}.

%%%%%%%%%%%%%%%%%%%%%%%%%%%%%%%%%%%%%%%%%%%%%%%%%%
%%%%%%%%%%%%%%%%%%%%%%%%%%%%%%%%%%%%%%%%%%%%%%%%%%
%%%%%%%%%%%%%%%%%%%%%%%%%%%%%%%%%%%%%%%%%%%%%%%%%%
%%%%%%%%%%%%%%%%%%%%%%%%%%%%%%%%%%%%%%%%%%%%%%%%%%
\section{Mass spectrum of hadrons}
\label{sec:had_spectrum}
%%%%%%%%%%%%%%%%%%%%%%%%%%%%%%%%%%%%%%%%%%%%%%%%%%
%%%%%%%%%%%%%%%%%%%%%%%%%%%%%%%%%%%%%%%%%%%%%%%%%%
%%%%%%%%%%%%%%%%%%%%%%%%%%%%%%%%%%%%%%%%%%%%%%%%%%
%%%%%%%%%%%%%%%%%%%%%%%%%%%%%%%%%%%%%%%%%%%%%%%%%%

In the confined phase of QCD, hadrons appear as a set of states characterized by their masses and quantum numbers. Neglecting finite-width effects, which is appropriate for the present purposes, the spectrum can be represented as a sum over discrete point-like ground-states and their resonances,
\begin{equation}
\label{eq:rho_pdg}
\rho(m)=\sum_i d_i\,\delta(m-m_i),
\end{equation}
where $m_i$ denotes the mass of a hadronic state and $d_i$ its degeneracy factor. 

It is often convenient to represent the spectrum with the cumulative distribution of states,
\begin{equation}
\label{eq:cum_def}
N(m)=\int\limits_0^{m}\mathrm{d}m'\,\rho(m'),
\end{equation}
which counts the total number of hadronic degrees of freedom with masses below $m$. For a spectrum composed of discrete resonances, Eq.~\eqref{eq:cum_def} reduces to
\begin{equation}
N(m)=\sum_i d_i\,\theta(m-m_i),
\end{equation}
where $\theta$ is the Heaviside step function.

Owing to the additive structure of the discrete density of states, the hadronic spectrum can be decomposed into mesonic and baryonic contributions, as well as into sectors carrying definite quantum numbers.

\begin{figure}[t!]
    \centering
    \includegraphics[width=1\linewidth]{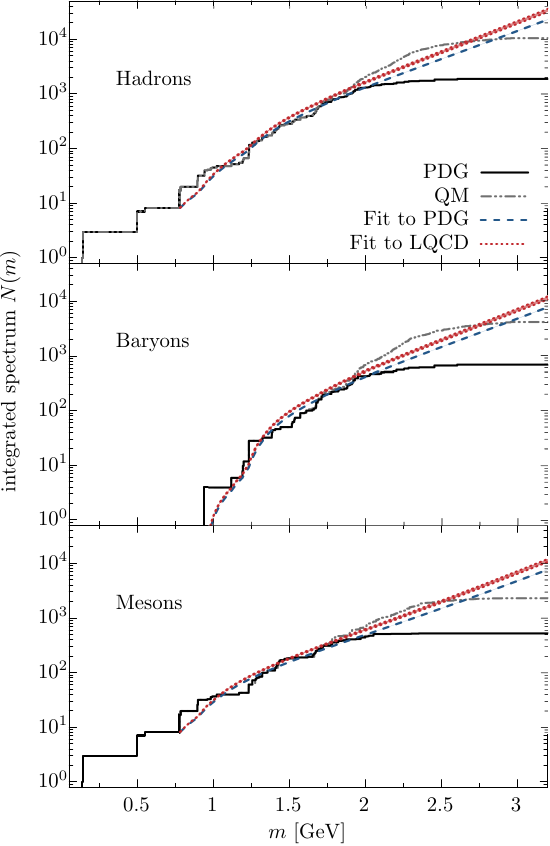}
    \caption{Cumulative spectra of hadrons (top panel), baryons (middle panel), and mesons (bottom panel) obtained from the PDG~\cite{ParticleDataGroup:2024cfk} (black solid lines). Also shown are spectra predicted by the quark model~\cite{Loring:2001ky, Ebert:2009ub} (gray dash-dotted lines). Continuous Hagedorn spectra are obtained from fits of the hadron spectrum to the PDG (blue dashed line) and lattice QCD (red dotted lines) results. The shaded bands indicate the uncertainty of the extracted Hagedorn limiting temperatures (see text for details). We note that the hadronic spectrum includes antibaryon contributions. The middle panel, however,  shows baryons only and does not include antibaryon contribution.}
    \label{fig:had_spec}
\end{figure}

In this work, we consider the hadronic spectrum of experimentally established states listed by the Particle Data Group (PDG) tables~\cite{ParticleDataGroup:2024cfk}. We include baryons with three- and four-star confidence ratings, as well as confirmed mesons. Broad scalar resonances such as $f_0(500)$ and $\kappa_0^\ast(700)$ are excluded due to the ambiguity associated with their interpretation as well-defined isolated states~\cite{Broniowski:2015oha, Friman:2015zua}. With this selection, the mesonic spectrum extends up to masses of approximately \mbox{$m \simeq2.4\,$GeV}, while established baryons reach masses of about \mbox{$m \simeq2.6\,$GeV}. Above these scales, the increasing widths, complicated decay properties, and overlap of resonances render a discrete description incomplete.

In Fig.~\ref{fig:had_spec}, we show the cumulative spectra of all hadrons, as well as of baryons and mesons separately. For masses up to \mbox{$m\lesssim 2\,$GeV}, the increase of the density of states of all hadrons is roughly linear on the logarithmic scale. A similar trend is seen in the spectra of baryons and mesons separately. We have also examined an extended hadronic spectrum that incorporates resonances predicted by the quark model (QM)~\cite{Loring:2001ky, Ebert:2009ub}. In this case, approximately linear behavior of the cumulative spectrum on a logarithmic scale persists up to masses of order $m\simeq 2.5\,$GeV. This observation supports an interpretation of the hadronic spectrum in terms of an exponential growth of states, consistent with a Hagedorn-type mass spectrum~\cite{Hagedorn:1965st, Hagedorn:1971mc}.

%%%%%%%%%%%%%%%%%%%%%%%%%%%%%%%%%%%%%%%%%%%%%%%%%%%%%%%
%%%%%%%%%%%%%%%%%%%%%%%%%%%%%%%%%%%%%%%%%%%%%%%%%%%%%%%
%%%%%%%%%%%%%%%%%%%%%%%%%%%%%%%%%%%%%%%%%%%%%%%%%%%%%%%
%%%%%%%%%%%%%%%%%%%%%%%%%%%%%%%%%%%%%%%%%%%%%%%%%%%%%%%
\section{String-based Hagedorn mass spectrum}
\label{sec:hagedorn}
%%%%%%%%%%%%%%%%%%%%%%%%%%%%%%%%%%%%%%%%%%%%%%%%%%%%%%%
%%%%%%%%%%%%%%%%%%%%%%%%%%%%%%%%%%%%%%%%%%%%%%%%%%%%%%%
%%%%%%%%%%%%%%%%%%%%%%%%%%%%%%%%%%%%%%%%%%%%%%%%%%%%%%%
%%%%%%%%%%%%%%%%%%%%%%%%%%%%%%%%%%%%%%%%%%%%%%%%%%%%%%%

An exponential growth of the hadronic mass spectrum seen in Fig.~\ref{fig:had_spec} is commonly associated with the existence of a limiting Hagedorn temperature $T_H$~\cite{Hagedorn:1965st, Hagedorn:1971mc} and can be parametrized by the exponential form, $\rho(m)\simeq m^ae^{m/T_H}$. While originally introduced within the statistical bootstrap framework, such a spectrum also arises naturally in relativistic string theory, where the proliferation of string excitation modes leads to an exponential density of states~\cite{Frautschi:1971ij, Green_Schwarz_Witten_2012}. In this setting, both the preexponential factor and the Hagedorn temperature $T_H$ are determined by the underlying string dynamics rather than being free parameters.

\begin{figure}[t!]
    \centering
    \includegraphics[width=1\linewidth]{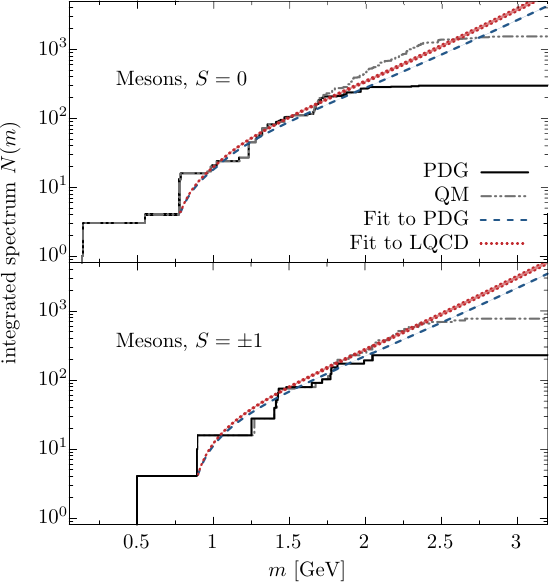}
     \caption{As in Fig.~\ref{fig:had_spec}, but for mesons separated into nonstrange (top panel) and strange (bottom panel) sectors.}
    \label{fig:mes_spec}
\end{figure}

\begin{figure}[t!]
    \centering
    \includegraphics[width=1\linewidth]{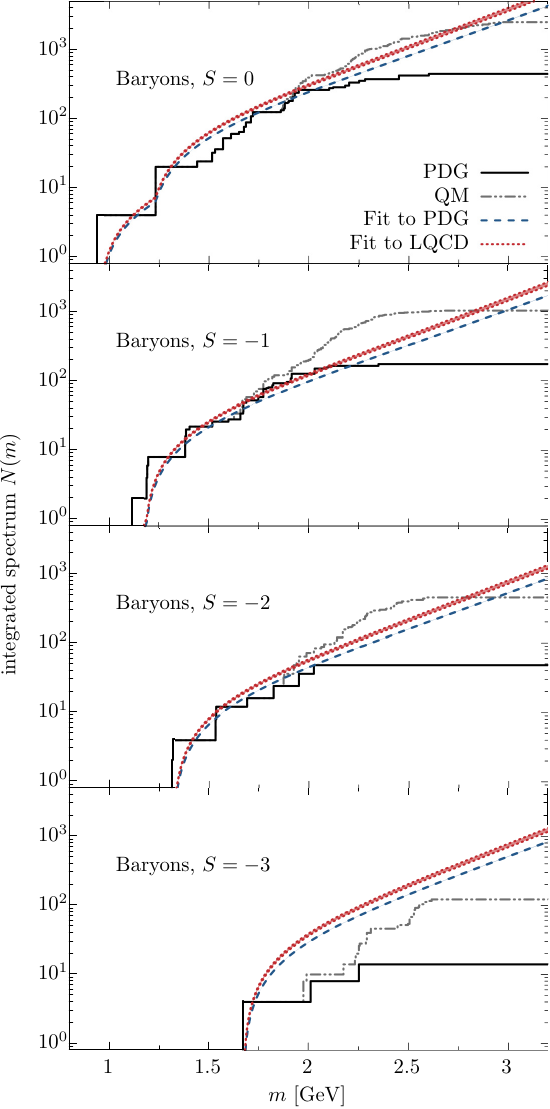}
    \caption{As in Fig.~\ref{fig:had_spec}, but for baryons separated into strangeness sectors $S=0,-1,-2,-3$, shown from top to bottom. We note that the spectra do not include antibaryons.}
    \label{fig:bar_spec}
\end{figure}

For a massless open string propagating in physical spatial dimension $d=3$, the Hagedorn temperature is directly related to the string tension $\sigma$, which sets the confinement scale. Using phenomenologically relevant values $\sqrt{\sigma}\simeq 450$--$500\,$MeV~\cite{Meyer:2004gx}, one obtains $T_H$ of the order of $300\,$MeV, comparable to the deconfinement temperature of pure gauge theory~\cite{Meyer:2009tq, Athenodorou:2020ani, Borsanyi:2022xml}. In addition to fixing the exponential growth rate, the string framework uniquely determines the power-law prefactor of the density of states, which depends on the dimensionality and on whether the string is open or closed~\cite{Green_Schwarz_Witten_2012}.

For an open relativistic string in four spacetime dimensions, the asymptotic density of states at large mass is given by
\begin{equation}
\label{eq:open_spec}
    \rho_{\rm str}(m)
    =
    \frac{\sqrt{2\pi}}{6\,T_H}
    \left(\frac{T_H}{m}\right)^{3/2}
    e^{m/T_H}\,,
\end{equation}
where $T_H$ denotes the Hagedorn temperature, which is related to the string tension $\sqrt\sigma$,
\begin{equation}
    T_H = \sqrt{\frac{3}{2\pi}} \sqrt{\sigma}.
\end{equation} The power-law prefactor reflects the two transverse degrees of freedom of the open string in four space-time dimensions, while the exponential factor encodes the rapid growth of the number of string excitation modes.

It is worth noting that Eq.~\eqref{eq:open_spec} represents the universal asymptotic density of states of an open string and is independent of the quantum numbers carried by the string endpoints. Flavor and spin enter only through endpoint degeneracies and physical threshold masses, while the string contribution to the spectrum remains universal. In the following, we employ this open-string density of states as the basic building block for describing mesonic and baryonic spectra. Mesons are identified with open strings carrying quark-antiquark endpoints~\cite{Fujimoto:2025sxx, Marczenko:2025nhj}, whereas baryons are described as open strings with quark-diquark endpoints~\cite{Fujimoto:2025trl}. In both cases, differences between hadron species arise solely from endpoint quantum numbers, while the string dynamics encoded in Eq.~\eqref{eq:open_spec} remains unchanged.

%%%%%%%%%%%%%%%%%%%%%%%%%%%%%%%%%%%%%%%%%%%%%%%%%%%%%%%
%%%%%%%%%%%%%%%%%%%%%%%%%%%%%%%%%%%%%%%%%%%%%%%%%%%%%%%
\subsection{Mesons}
%%%%%%%%%%%%%%%%%%%%%%%%%%%%%%%%%%%%%%%%%%%%%%%%%%%%%%%
%%%%%%%%%%%%%%%%%%%%%%%%%%%%%%%%%%%%%%%%%%%%%%%%%%%%%%%

We model the continuous Hagedorn spectrum of mesons as introduced in Refs.~\cite{Fujimoto:2025sxx, Marczenko:2025nhj, Fujimoto:2025trl}. A common practice is to take out the Nambu-Goldstone mesons, i.e., $\lbrace\pi,~K,~\eta\rbrace$, from the continuous part of the spectrum and treat them as point-like states. Effectively, a low-mass threshold for the continuous part is introduced. The meson mass spectrum then reads
\begin{equation}\label{eq:rho_mes}
    \rho_{\rm mes}(m)
    =
    \sum_{i = \pi, K,\eta} d_i\delta\left(m - m_i\right)+ d_{\rm mes}(m) \rho_{\rm str}(m),
\end{equation}
The discrete part of the spectrum includes the Nambu-Goldstone mesons with degeneracy $d_i$. The degeneracy factor $d_{\rm mes}(m)$ in the continuous part depends on the strangeness content of the string configuration and accounts for the threshold masses:
\begin{equation}\label{eq:d_mes}
    d_{\rm mes}(m) = \sum_i d_i \theta\left(m-m_i\right),
\end{equation}
where $\theta$ is the Heaviside step function. The degeneracies $d_i$'s and $m_i$'s, and quantum numbers for mesons are listed in Table~\ref{tab:mes}.

\begin{table}[t!]
    \centering
    \begin{tabular}{c||cccccc}
    \hline
        Channel                    & mass~[GeV] & $I$ & $Q$ & $S$    & deg \\\hline
        $\rho$ ($l\bar l$)          & 0.770   &   1 &  $0,\pm 1$   &  0     &  16 \\
        $K^\ast$ ($l\bar s,\,s\bar l$) & 0.896   & 1/2 &  $0,\pm 1$   & $\pm 1$ &  16 \\
        $\phi$   ($s\bar s$)          & 1.019   &   0 &   0  &      0 &   4 \\
        \hline
    \end{tabular}
    \caption{Parameters of the degeneracy factor $d_{\rm mes}(m)$ in Eq.~\eqref{eq:d_mes} for mesons. Here $l=u,d$ denotes a light quark, $s$ a strange quark, and a bar indicates antiquarks.}
    \label{tab:mes}
\end{table}

%%%%%%%%%%%%%%%%%%%%%%%%%%%%%%%%%%%%%%%%%%%%%%%%%%%%%%%
%%%%%%%%%%%%%%%%%%%%%%%%%%%%%%%%%%%%%%%%%%%%%%%%%%%%%%%
\subsection{Baryons}
%%%%%%%%%%%%%%%%%%%%%%%%%%%%%%%%%%%%%%%%%%%%%%%%%%%%%%%
%%%%%%%%%%%%%%%%%%%%%%%%%%%%%%%%%%%%%%%%%%%%%%%%%%%%%%%

\begin{table}[t!]
    \centering
    \begin{tabular}{c||ccccc}
    \hline
        Channel   & mass~[GeV] & $I$   & $Q$            & $S$ & deg \\\hline
        $N$       ($l[ll]$) & 0.938  & $1/2$ & $(0,~1)$       &  0 & 4    \\
        $\Lambda$ ($s[ll]$) & 1.116  & $0$   & $(0)$          & -1 & 2    \\
        $\Sigma$  ($l[ls]$) & 1.189  & $1$   & $(0,~\pm 1)$   & -1 & 6    \\
        $\Delta$  ($l\lbrace ll\rbrace$) & 1.232  & $3/2$ & $(0,~\pm 1,~2)$&  0 & 16   \\
        $\Xi$     ($s[ls]$) & 1.315  & $1/2$ & $(0,~-1)$      & -2 & 4    \\
        $\Omega$  ($s\lbrace ss\rbrace$) & 1.672  & $0$   & $(-1) $        & -3 & 4    \\
        \hline
    \end{tabular}
    \caption{Parameters of the degeneracy factor $d_{\rm bar}(m)$ in Eq.~\eqref{eq:d_bar} for baryons. Diquarks are indicated by brackets, with $\lbrace xx \rbrace$ and $[xx]$ denoting symmetric (bad diquark) and antisymmetric (good diquark) flavor configurations, respectively (see Ref.~\cite{Fujimoto:2025trl} for details). Here $l=u,d$ denotes a light quark, $s$ a strange quark. We note that the degeneracies do not account for the contribution of antibaryons.}
    \label{tab:bar}
\end{table}

It was recently demonstrated that the baryon spectrum can be described by modeling baryons as open relativistic strings with a quark attached to one endpoint and a diquark attached to the other~\cite{Fujimoto:2025trl}. In this description, the string dynamics and the resulting excitation spectrum are identical to those governing mesons, such that the asymptotic density of states is given by the same open-string expression. Baryon-specific properties enter only through the physical threshold masses and the degeneracy factors associated with the quantum numbers of the string endpoints.

The use of diquark degrees of freedom is motivated by the color structure of QCD. Two quarks combine in color space as $\mathbf{3}\otimes\mathbf{3}=\bar{\mathbf{3}}\oplus\mathbf{6}$, where the antisymmetric color antitriplet channel is attractive, while the symmetric sextet channel is repulsive. As a consequence, correlated quark pairs in the color-$\bar{\mathbf{3}}$ channel provide a natural effective degree of freedom inside baryons and are treated here as effective string endpoints. Following this construction, the baryonic mass spectrum is written as
\begin{equation}
\label{eq:rho_bar}
    \rho_{\rm bar}(m) = d_{\rm bar}(m)\,\rho_{\rm str}(m),
\end{equation}
where $\rho_{\rm str}(m)$ denotes the universal open-string density of states. The degeneracy factor $d_{\rm bar}(m)$ encodes the endpoint quantum numbers and the physical thresholds associated with different strangeness sectors and is given by
\begin{equation}
\label{eq:d_bar}
    d_{\rm bar}(m) = \sum_i d_i\,\theta\!\left(m-m_i\right).
\end{equation}
The values of the threshold masses $m_i$, the corresponding degeneracy factors $d_i$, and quantum numbers are summarized in Table~\ref{tab:bar}. We note that the tabulated degeneracies do not account for the contribution of antibaryons.

%%%%%%%%%%%%%%%%%%%%%%%%%%%%%%%%%%%%%%%%%%%%%%%%%%%%%%%
%%%%%%%%%%%%%%%%%%%%%%%%%%%%%%%%%%%%%%%%%%%%%%%%%%%%%%%
\subsection{Hadrons}
%%%%%%%%%%%%%%%%%%%%%%%%%%%%%%%%%%%%%%%%%%%%%%%%%%%%%%%
%%%%%%%%%%%%%%%%%%%%%%%%%%%%%%%%%%%%%%%%%%%%%%%%%%%%%%%

The continuous mesonic and baryonic spectra are constructed independently and can be combined to form the full hadronic spectrum:
\begin{equation}\label{eq:had_spec}
    \rho_{\rm had}(m) = \rho_{\rm mes}(m) + 2\rho_{\rm bar}(m),
\end{equation}
where the factor $2$ in the baryonic spectrum accounts for the contribution of antibaryons. 

The analysis of the experimental hadron spectrum, in the context of Hagedorn exponential form $\rho(m) \simeq m^a e^{m/T_H}$, has been extensively discussed in the literature~\cite{Hagedorn:1971mc, Rafelski:2016nxx, Majumder:2010ik, Lo:2015cca, ManLo:2016pgd, Broniowski:2000bj, Broniowski:2004yh}. The extracted value of the Hagedorn temperature in such analyses is model-dependent and sensitive to the assumed preexponential factor, typically yielding values below $200~$MeV.

To determine the Hagedorn temperature, we perform a non-linear least-squares fit of the hadronic spectrum in Eq.~\eqref{eq:had_spec} to the discrete cumulative spectrum of all hadrons from the PDG in the mass range $0.77 < m < 2\,\mathrm{GeV}$, where the exponential trend is clearly visible. This approach complements the analysis of Ref.~\cite{Fujimoto:2025trl}, where the mesonic and baryonic spectra were treated separately. Since the experimentally established hadronic spectrum provides the most complete and reliable input, we consider a combined fit to all hadrons to be the most robust determination of $T_H$. The resulting value is $T_H = 340.1 \pm 0.3\,\mathrm{MeV}$. For completeness, we also perform separate fits to the baryonic and mesonic spectra in the mass ranges $1 < m \lesssim 2\,\mathrm{GeV}$ and $0.77 < m \lesssim 2\,\mathrm{GeV}$, respectively. In both cases, we obtain very similar values, $T_H \simeq 340\,\mathrm{MeV}$, thereby demonstrating that a single Hagedorn temperature provides a consistent description of both mesonic and baryonic spectra, in line with the present string-based description. We note that the value of $T_H$ is directly proportional to the string tension $\sqrt\sigma$. The recent estimates, $\sqrt\sigma = 0.485(6)~$GeV~\cite{Athenodorou:2020ani} and $\sqrt\sigma = 0.4817(97)\,$GeV~\cite{Brambilla:2022het}, give $T_H=0.335(4)~$GeV and $T_H=0.3329(67)~$GeV, respectively.

\begin{figure}[t!]
    \centering
    \includegraphics[width=1\linewidth]{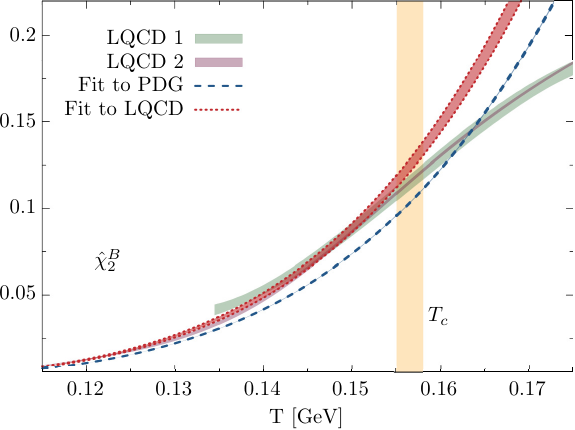}
    \caption{The second-order fluctuations of the  net-baryon number, $\hat \chi_2^B$. Shown are the results obtained from the fit of the continuous to all hadrons in the PDG spectrum (blue dashed line) and the fit to the lattice QCD results (red dotted line). The bands represent the uncertainty in the obtained Hagedorn limiting temperatures (see text for details). We note that the uncertainty associated with the PDG-based fit is smaller than the line width on the scale shown. The LQCD results are taken from Ref.~\cite{Bollweg:2021vqf} (LQCD 1) and~\cite{Abuali:2025tbd} (LQCD 2). They correspond to continuum extrapolated and continuum estimated results, respectively. The vertical yellow band marks the temperature of the chiral crossover $T_c = 156.5\pm1.5~$MeV~\cite{HotQCD:2018pds}.}
    \label{fig:xbb}
\end{figure}

We compare the obtained continuous Hagedorn spectra for hadrons, as well as baryons and mesons separately, with the discrete spectra from PDG and the quark model in Fig.~\ref{fig:had_spec}. The continuous spectrum fitted to the PDG data provides an excellent description of the experimentally established states over the fitted mass range. In contrast, the QM cumulative spectra exhibit a noticeably steeper growth with mass above $2\,$GeV, indicating a higher density of states at large masses. As a result, a continuous spectrum constrained by the PDG data underestimates the number of states predicted by the quark model at higher masses.

Once the Hagedorn temperature is fixed, the string-based construction determines the continuous mass spectra in individual strangeness sectors for both mesons and baryons. Figure~\ref{fig:mes_spec} shows the cumulative spectra of nonstrange and strange mesons. The continuous spectrum constrained by the PDG provides a good description of the experimentally established states in both sectors. The definite-strangeness baryonic spectra are displayed in Fig.~\ref{fig:bar_spec}. For nonstrange and singly strange baryons, the continuous spectrum fitted to the PDG data closely follows the experimentally observed cumulative spectrum over the accessible mass range. In the $S=-2$ sector, the Hagedorn spectrum slightly overestimates the PDG spectrum. In the $S=-3$ sector, the extracted continuous Hagedorn spectrum predicts a significantly larger density of states than is currently established experimentally. However, owing to the large threshold mass, this excess in the continuous spectrum has a negligible impact on thermodynamic observables in the temperature range considered~\cite{ManLo:2016pgd}.

%%%%%%%%%%%%%%%%%%%%%%%%%%%%%%%%%%%%%%%%%%%%%%%%%%
%%%%%%%%%%%%%%%%%%%%%%%%%%%%%%%%%%%%%%%%%%%%%%%%%%
%%%%%%%%%%%%%%%%%%%%%%%%%%%%%%%%%%%%%%%%%%%%%%%%%%
%%%%%%%%%%%%%%%%%%%%%%%%%%%%%%%%%%%%%%%%%%%%%%%%%%
\section{Fluctuations of conserved charges}
\label{sec:fluct}
%%%%%%%%%%%%%%%%%%%%%%%%%%%%%%%%%%%%%%%%%%%%%%%%%%
%%%%%%%%%%%%%%%%%%%%%%%%%%%%%%%%%%%%%%%%%%%%%%%%%%
%%%%%%%%%%%%%%%%%%%%%%%%%%%%%%%%%%%%%%%%%%%%%%%%%%
%%%%%%%%%%%%%%%%%%%%%%%%%%%%%%%%%%%%%%%%%%%%%%%%%%

\begin{figure}[t!]
    \centering
    \includegraphics[width=1\linewidth]{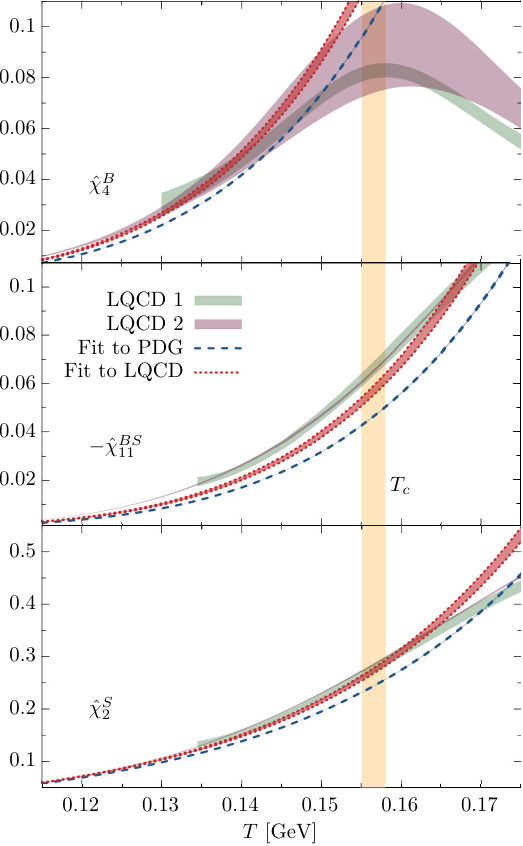}
    \caption{As in Fig.~\ref{fig:xbb} but for the fourth-order fluctuations of the net-baryon number, $\hat \chi_4^B$ (top panel), baryon-strangeness correlator, $\hat \chi_{11}^{BS}$ (middle panel), and the second-order fluctuations of strangeness, $\hat \chi_2^S$ (bottom panel).}
    \label{fig:panel_b}
\end{figure}

We now turn to fluctuations and correlations of conserved charges, which provide a sensitive probe of the hadronic spectrum entering the thermodynamics in different sectors of quantum numbers~\cite{Bazavov:2014xya, Lo:2015cca, ManLo:2016pgd, Bollweg:2021vqf}. In particular, the net-baryon number fluctuations are directly controlled by the baryonic sector and therefore offer a stringent test of the string-based spectrum constructed in the previous section. 

To investigate the thermodynamic behavior of the hadronic phase, we model the system as a mixture of ideal gases of hadronic states with an exponential Hagedorn mass spectrum. 

At finite temperature $T$ and chemical potential $\mu$, the normalized pressure $\hat p = p/T^4$ of a noninteracting relativistic particle species with mass $m$ in the Boltzmann approximation per degree of freedom reads
\begin{equation}\label{eq:pressure}
    \hat p(T, \mu, m) = \frac{1}{2\pi^2}\frac{m^2}{T^2} K_2\left(\frac{m}{T}\right)e^{\mu/T},
\end{equation}
where $K_2$ is the modified Bessel function of the second kind. For a particle carrying baryon number $B$, strangeness $S$ and electric charge $Q$, the chemical potential reads
\begin{equation}
    \mu = B\mu_B + Q\mu_Q + S\mu_S,
\end{equation}
where $\mu_{B/Q/S}$ are the chemical potentials associated with quantum numbers. 

Given the density of states $\rho(m)$, the total pressure can be written as
\begin{equation}
    \hat p(T, \mu) = \int \dd m\; \rho(m) \hat p(T,\mu,m).
\end{equation}

The fluctuations of conserved charges are obtained through generalized susceptibilities~\cite{Allton:2005gk, Karsch:2010ck}. The susceptibility of order $n$ is given as a derivative of the thermodynamic pressure with respect to associated chemical potentials,
\begin{equation}
    \hat \chi^{\rm BQS}_{ijk}(T,\mu) = \frac{\pp^n \hat p(T, \mu) }{\pp \hat\mu_B^i\pp\hat\mu_Q^j\pp\hat\mu_S^k}\Bigg|_T,
    \quad\quad i+j+k = n,
\end{equation}
where $\hat \mu = \mu / T$.

We mostly focus on the second-order fluctuations and correlations of conserved charges quantified by the $n=2$ generalized susceptibilities. We take advantage of the fact that in the O(4) universality class, the $n=2$ susceptibilities are non-singular in the chiral limit~\cite{Friman:2011pf}. Thus, they are dominantly quantified by the regular part of the thermodynamic potential, which we assume can be approximated by the thermodynamic pressure introduced in Eq.~\eqref{eq:pressure}. In the context of probability distributions, this is equivalent to the assumption that Skellam-like functions~\cite{Braun-Munzinger:2011xux} are a good approximation of the width of the QCD net-charge distributions in the hadronic phase. The higher-order susceptibilities are sensitive to the tail of distributions, which are affected by a singular scaling function and non-resonant interactions in the hadronic phase, thus deviating from predictions of a Skellam-like function.

Using the continuous spectrum constrained by the PDG input, we evaluate the second-order net-baryon number susceptibility, $\hat\chi_2^{B}$. The result, shown in Fig.~\ref{fig:xbb}, lies systematically below the LQCD data up to the chiral crossover region. Such a deviation is commonly interpreted as evidence that the experimental PDG spectrum is incomplete, particularly in the baryonic sector~\cite{Majumder:2010ik, Lo:2015cca, ManLo:2016pgd}.

For illustration, we also consider the fourth-order cumulant $\hat\chi_4^B$, which provides a more sensitive probe of the tail of the net-baryon probability distribution, which is not fully captured within an ideal-gas description. As shown in the top panel of Fig.~\ref{fig:panel_b}, the underestimation of fluctuations in $\hat\chi_2^B$ persists at the fourth-order susceptibility. We note, however, that the thermodynamic pressure in Eq.~\eqref{eq:pressure}, which leads to the Skellam distribution of net-baryon number fluctuations, implies that $\hat\chi_4^B/\hat\chi_2^B=1$ for any $\rho(m)$ and temperature. Thus, it is not possible to simultaneously improve $\hat\chi_2^B$ and $ \hat\chi_4^B$ by modeling $\rho(m)$ as LQCD results show   $\hat\chi_4^B/\hat\chi_2^B< 1$ towards the chiral crossover~\cite{Bazavov:2017dus}.

Having established the behavior of net-baryon number fluctuations, we now turn to observables involving strangeness. Figure~\ref{fig:panel_b} shows the second-order baryon-strangeness correlator $\hat \chi_{11}^{BS}$ and the strangeness fluctuation $\hat \chi_2^{S}$. We note that $\hat\chi_{11}^{BS}$ is dominated by contributions associated with singly strange baryons, while the influence of doubly and triply strange states is strongly suppressed in the temperature range shown~\cite{Lo:2015cca}. This is because higher strangeness sectors are characterized by substantially larger threshold masses. On the other hand, the strangeness fluctuation $\hat\chi_2^{S}$,  receives contributions from both strange mesons and strange baryons. At lower temperatures, this observable is dominated by the contribution from mesons, while baryons become increasingly important toward the crossover region.

Comparing the model and  LQCD results in Fig.~\ref{fig:panel_b} we find that the PDG-constrained spectrum in the strangeness sector is also insufficient to explain either of the observables. This is consistent with earlier findings based on a parameterized Hagedorn spectrum that suggested the existence of uncharted strange hadrons in the intermediate mass range~\cite{Lo:2015cca, ManLo:2016pgd}.

The evaluation of electric-charge fluctuations and mixed correlators requires an additional step. In the string-based description, degeneracy factors arise from summing over the spin and flavor degrees of freedom associated with the endpoints of an open string. In particular, the compact degeneracy factors introduced in Refs.~\cite{Fujimoto:2025sxx} (summarized in Tables~\ref{tab:mes} and~\ref{tab:bar}) implicitly include all members of a given isospin multiplet. While this is sufficient for bulk thermodynamic quantities, a resolution into individual isospin components is necessary for observables that depend explicitly on electric charge. Accordingly, each flavor channel is decomposed into its isospin structure, and each member of the multiplet is treated as a distinct state carrying definite quantum numbers. Neglecting minor strong isospin breaking effects, we assume that all isospin components have the same mass and are equally Boltzmann suppressed. With this prescription, the contribution of a mesonic or baryonic channel to thermodynamic observables can be written as
\begin{equation}
d_{\mathrm{mes}/\mathrm{bar}}(m)
=
\sum_i d_i \theta \left(m-m_i\right)\sum_{k=1}^{d_i^{I}} \frac{1}{d_i^{I}},
\end{equation}
where $i$ labels the isospin multiplet, $k$ enumerates its members, and $d_i^{I}=2I_i+1$ denotes the isospin degeneracy of multiplet $i$. This construction preserves the total degeneracy of each hadronic channel while enabling an unambiguous assignment of conserved quantum numbers to individual states. This allows for a consistent evaluation of charge-dependent susceptibilities and mixed correlators such as $\hat \chi_2^{Q}$, $\hat \chi_{11}^{BQ}$, and $\hat \chi_2^{QS}$.

\begin{figure}[t!]
    \centering
    \includegraphics[width=1\linewidth]{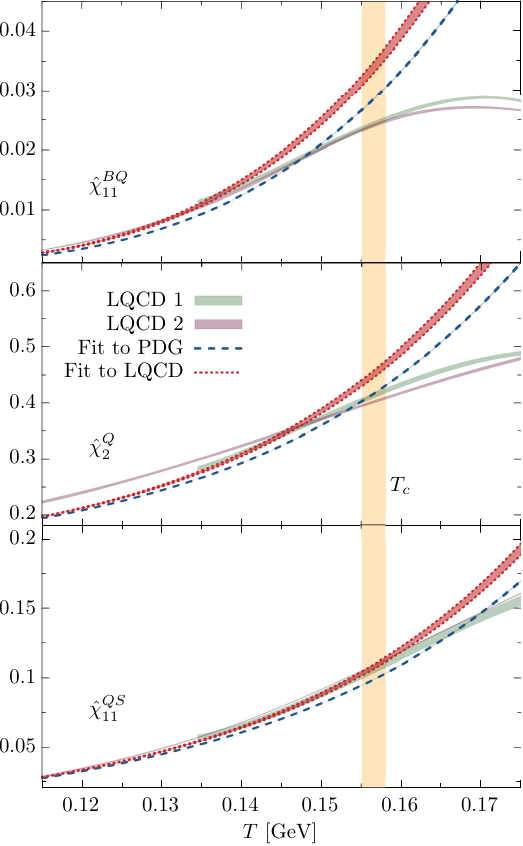}
    \caption{As in Fig.~\ref{fig:xbb} but for the baryon-charge correlator, $\hat \chi_{11}^{BQ}$ (top panel), second-order electric-charge fluctuations, $\hat \chi_{2}^{Q}$ (middle panel), and the charge-strangeness correlator, $\hat \chi_{11}^{QS}$ (bottom panel).}
    \label{fig:panel_q}
\end{figure}

The comparison with electric-charge fluctuations obtained in lattice QCD simulations is shown in Fig.~\ref{fig:panel_q}. We find that, in the charge sector, the string spectrum fitted to the PDG states systematically underestimates fluctuations in the temperature range approaching the crossover.

The overall comparison with LQCD observables demonstrates that a continuous Hagedorn spectrum constrained by experimentally established PDG states is not sufficient to provide a quantitative description of the second-order fluctuations and correlations of conserved charges, in particular in the vicinity of the chiral crossover. The systematic underestimation of fluctuations of conserved charges indicates that the effective hadronic spectrum relevant for thermodynamics differs from that inferred from spectroscopy alone. This motivates an alternative determination of the Hagedorn temperature directly from lattice QCD thermodynamics. 

\begin{figure}[t!]
    \centering
    \includegraphics[width=1\linewidth]{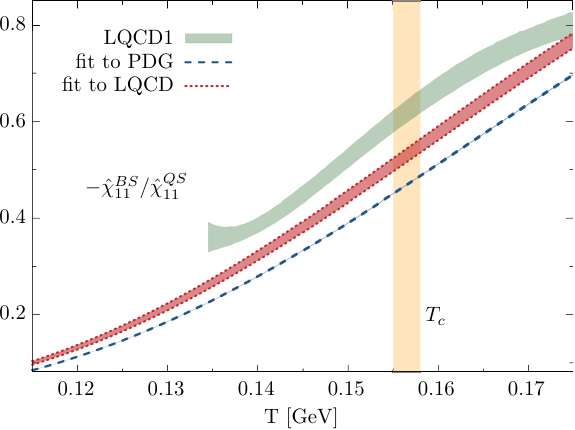}
    \caption{As in Fig.~\ref{fig:xbb} but for the ratio of baryon-strangeness and charge-strangeness correlator, $-\hat \chi_{11}^{BS} / \hat \chi_{11}^{QS}$.}
    \label{fig:chi_bs_qs}
\end{figure}

We therefore extract the Hagedorn temperature directly from lattice QCD by fitting the string-based baryonic spectrum to the second-order net-baryon number susceptibility in the temperature range $T \leq 155\,$MeV. This procedure yields a Hagedorn temperature \mbox{$T_H = 323(3)\,$MeV}. The obtained spectra, shown in Figs.~\ref{fig:had_spec},~\ref{fig:mes_spec}, and~\ref{fig:bar_spec}, display a systematic increase of states as compared to the PDG-induced fit due to smaller $T_H$. The resulting spectrum leads to a markedly improved description of lattice QCD results for a broad set of fluctuations and correlations of conserved charges, as shown in Figs.~\ref{fig:xbb}, \ref{fig:panel_b}, and~\ref{fig:panel_q}. All observables shown display better agreement with LQCD results, except $\hat \chi_{11}^{BS}$ and $\hat \chi_{11}^{BQ}$, as also seen in Fig.~\ref{fig:chi_bs_qs} in the  $\chi^{BS}_{11}/\hat \chi^{QS}_{11}$ ratio.

The obtained baryon-strangeness correlator underestimates the LQCD results. This is partly because the Hagedorn temperature was extracted from $\hat \chi_2^{B}$, which is mostly driven by nonstrange baryons and is not very sensitive to the strange baryons. This may point to the existence of unconfirmed or hitherto unknown strange baryons in accordance with previous findings~\cite{Lo:2015cca, ManLo:2016pgd}. Indeed, from Figs.~\ref{fig:panel_b} and~\ref{fig:chi_bs_qs} one sees that the string mass spectrum increases $\hat\chi_{BS}$ towards the LQCD results as compared to the PDG mass spectrum.

Furthermore, in the strange-baryon sector, the improvement of interactions within the S-matrix formalism was shown to increase further the strange-baryon correlations towards the LQCD results~\cite{Fernandez-Ramirez:2018vzu}.

On the other hand, the baryon-charge correlator overshoots the LQCD result. In general, $\hat \chi_{11}^{BQ}$ is dominated by contributions from light, nonstrange baryons. However, the ideal-gas treatment used in this work neglects repulsive and nonresonant interactions. A consistent implementation of these effects in the S-matrix approach leads to a substantial reduction of $\hat \chi_{11}^{BQ}$ towards chiral crossover~\cite{Andronic:2018qqt}. Since the present Hagedorn approach encodes interactions only implicitly through an effective density of states, the remaining overestimation of $\hat\chi_{11}^{BQ}$ indicates that interaction-induced repulsion in the light baryon sector is not captured, despite the improved spectral description. 

Overall, the Hagedorn spectrum extracted directly from lattice QCD provides a consistent and quantitatively improved description of fluctuations of conserved charges, supporting a quark-diquark picture of baryons within a string-based framework. Recently, it was shown that such a description can also be successfully extended to the heavy-quark sector~\cite{Marczenko:2026drt}.

%%%%%%%%%%%%%%%%%%%%%%%%%%%%%%%%%%%%%%%%%%%%%%%%%%
%%%%%%%%%%%%%%%%%%%%%%%%%%%%%%%%%%%%%%%%%%%%%%%%%%
%%%%%%%%%%%%%%%%%%%%%%%%%%%%%%%%%%%%%%%%%%%%%%%%%%
%%%%%%%%%%%%%%%%%%%%%%%%%%%%%%%%%%%%%%%%%%%%%%%%%%
\section{Conclusions}
\label{sec:conclusions}
%%%%%%%%%%%%%%%%%%%%%%%%%%%%%%%%%%%%%%%%%%%%%%%%%%
%%%%%%%%%%%%%%%%%%%%%%%%%%%%%%%%%%%%%%%%%%%%%%%%%%
%%%%%%%%%%%%%%%%%%%%%%%%%%%%%%%%%%%%%%%%%%%%%%%%%%
%%%%%%%%%%%%%%%%%%%%%%%%%%%%%%%%%%%%%%%%%%%%%%%%%%

In this work, we examined the role of the hadronic mass spectrum in determining fluctuations and correlations of conserved charges in QCD. Our analysis was based on a string-inspired description of hadrons, in which mesons and baryons are modeled as open relativistic strings with quark-antiquark and quark-diquark endpoints, respectively. Within this framework, the exponential growth of the density of states is controlled by a Hagedorn temperature directly related to the string tension with a precisely derived prefactor.

We showed that continuous Hagedorn spectra constrained by the experimentally established hadron spectrum systematically underestimate fluctuations and correlations of conserved charges in the vicinity of the chiral crossover. This indicates that the hadronic spectrum relevant for thermodynamics is not fully captured by spectroscopy alone. Motivated by this observation, we determined the effective Hagedorn temperature directly from lattice QCD by fitting the string-based hadronic spectrum to the second-order net-baryon number susceptibility. This procedure yields a Hagedorn temperature $T_H = 323(3)\,$MeV.

% $T_H \simeq 323\,$MeV.

The Hagedorn spectrum extracted in this way leads to a markedly improved description of a broad set of fluctuations and correlations of conserved charges computed in lattice QCD, without introducing additional free parameters. Residual deviations observed in selected correlators can be traced to additional interaction effects, rather than to deficiencies of the spectral construction itself. Therefore, the resulting agreement with lattice QCD provides thermodynamic support for a string-based quark-diquark picture of baryons in the confined phase of QCD.

%%%%%%%%%%%%%%%%%%%%%%%%%%%%%%%%%%%%%%%%%%%%%%%%%%
%%%%%%%%%%%%%%%%%%%%%%%%%%%%%%%%%%%%%%%%%%%%%%%%%%
%%%%%%%%%%%%%%%%%%%%%%%%%%%%%%%%%%%%%%%%%%%%%%%%%%
%%%%%%%%%%%%%%%%%%%%%%%%%%%%%%%%%%%%%%%%%%%%%%%%%%
\medskip
\section*{Acknowledgments}
%%%%%%%%%%%%%%%%%%%%%%%%%%%%%%%%%%%%%%%%%%%%%%%%%%
%%%%%%%%%%%%%%%%%%%%%%%%%%%%%%%%%%%%%%%%%%%%%%%%%%
%%%%%%%%%%%%%%%%%%%%%%%%%%%%%%%%%%%%%%%%%%%%%%%%%%
%%%%%%%%%%%%%%%%%%%%%%%%%%%%%%%%%%%%%%%%%%%%%%%%%%
The authors acknowledge helpful discussions with Larry McLerran and Yuki Fujimoto, Chihiro Sasaki, and Gy\H{o}z\H{o} Kov\'acs. K.R. acknowledges discussions with Frithjof Karsch and support from the National Science Centre (NCN), Poland, under OPUS Grant No. 2022/45/B/ST2/01527. M. M. is supported by the World Premier International Research Center Initiative (WPI) under MEXT, Japan. K.R. also acknowledges the support of the Polish Ministry of Science and Higher Education. 
%%%%%%%%%%%%%%%%%%%%%%%%%%%%%%%%%%%%%%%%%%%%%%%%%%
%%%%%%%%%%%%%%%%%%%%%%%%%%%%%%%%%%%%%%%%%%%%%%%%%%
%%%%%%%%%%%%%%%%%%%%%%%%%%%%%%%%%%%%%%%%%%%%%%%%%%
%%%%%%%%%%%%%%%%%%%%%%%%%%%%%%%%%%%%%%%%%%%%%%%%%%

\bibliography{biblio}

\end{document}